\documentclass[11pt,twoside]{article}

\usepackage{asp2006}
\usepackage{epsf}
\usepackage{graphicx}
\usepackage{lscape}

\begin{document}
\title{The origin of star formation since z=1}   
\author{F. Hammer,\altaffilmark{1} M. Puech,\altaffilmark{2} Y. B. Yang,\altaffilmark{1} and H. Flores \altaffilmark{1}}   
\affil{\altaffilmark{1} GEPI, Observatoire de Paris, CNRS, Universite Paris Diderot, 92195 Meudon, France}    
\affil{\altaffilmark{2} ESO, Karl-Schwarzschild-Str. 2, D-85748 Garching bei Munchen, Germany}

\begin{abstract} 
The use of multiple integral field units with FLAMES/GIRAFFE at VLT
has revolutionized investigations of distant galaxy kinematics. This
facility may recover the velocity fields of almost all emission line
galaxies with $I$(AB)$< $22.5 at z$<$0.8. We have gathered a unique
sample of 63 velocity fields at $z\!=\!0.4$\,--\,0.75, which are
representative of $M_{stellar}\!\ge\! 1.5\!\times\!10^{10}M_{\odot}$
emission line ($W_{0}([OII])\ge15$\AA)galaxies, and are unaffected by
cosmic variance. Taking into account all galaxies -with or without
emission lines- in that redshift range, we find that 42$\,\pm\,$7\% of
them have anomalous kinematics, including 26$\,\pm\,$7\% with complex
kinematics, i.e. not supported by either rotation or by dispersion.
The large fraction of complex velocity fields suggests a large impact
of merging in shaping the galaxies in the intermediate mass range.
We discuss how this can be accommodated within the frame of current
scenarios of galaxy formation, including for the Milky Way and M31.
\end{abstract}


\section{Introduction}   
The cosmic star formation rate has declined by a factor $\sim\!10$
since $z\!\sim\!1$ (Lilly et al.\ 1996; Flores et al.\ 1999). At z=1,
the cosmic star formation density is clearly dominated by luminous
infrared galaxies (LIRGs, Flores et al.\ 1999; Le Flo'ch et al, 2006),
which number density evolves extremely rapidly. There is an
overall agreement between the integrated star formation rate (derived
from mid-IR observations from ISO then Spitzer) and the stellar mass
evolution (derived from near-IR observations). One may focus on the
evolving properties of intermediate mass galaxies (or $M*$ galaxies)
with stellar masses from 3$\!\times\!10^{10}$ to
3$\!\times\!10^{11}M_{\odot}$: they include most (from 65 to 80\%) of
the present-day stellar mass. About half of their stellar mass was
formed since z=1, mostly during strong star formation episodes during
which galaxies take the appearance of luminous infrared galaxies
(LIRGs, Hammer et al.\ 2005).

What were the physical processes driving star-formation at these
redshifts? Internal kinematics of distant galaxies is a powerful tracer of
the major processes governing star-formation and galaxy evolution in
the early universe such as merging, accretion, and hydrodynamic
feedback related to star-formation and AGN (e.g., Barnes \& Hernquist
1996; Barton et al.\ 2000; Dressler 2004). Thus, robustly measuring
the internal kinematics of distant galaxies is
crucial for understanding how galaxies formed and
evolved.

Flores et al.\ (2006) have presented the first study of a small sample
of 35 I(AB)$<$22.5 galaxies at z=0.4-0.75 using the integral-field
multi-object spectrograph GIRAFFE on the ESO-VLT. They found a large
fraction of galaxies with anomalous velocity fields, i.e. with
velocity fields discrepant from those of rotationally or dispersion (pressure)
supported galaxies. Interestingly they found that the large dispersion
of the Tully-Fisher at moderate z (see e.g. Conselice et al, 2005) is
mainly caused by galaxies with anomalous kinematics.

Here we present the kinematics of a representative sample of 63 $M*$ galaxies at z=0.4-0.75 (section 2), we discuss their properties (section 3) and relate them to  local galaxies, including the
Milky Way and M31 (section 4).

\section{Velocity fields of 63 $M*$ galaxies at z=0.4-0.75} 
\subsection{A representative sample of distant galaxies}
A Large Program at VLT entitled IMAGES (Intermediate MAss Galaxy
Evolution Sequence) is aiming to derive velocity fields from
VLT/GIRAFFE for galaxies in the Chandra Deep Field (CDFS) and combine
these observations with deep and high quality images from HST/ACS as
well as with deep mid-IR photometry from SPITZER/MIPS. It has now
reached more than half its completion and recently 39 velocity fields
of galaxies have been derived (Yang et al., 2007). Combined with the
previous observations of Flores et al. (2006), it leads to a sample of
74 galaxies. Observing time with VLT/GIRAFFE was ranging from 8 to 13
hours in order to provide high S/N in the 20 GIRAFFE-IFU pixels for
each galaxy. In most cases the [OII] emission doublet is well resolved
at a resolution of 9000$\times$(1+z). It provides us with 68 robust
velocity fields for which there are 6 to 18 GIRAFFE pixels with high
S/N. The remaining 6 galaxies have been rejected from the sample
because their velocity fields include less than 4 pixels with S/N$>$4
(see e.g. Flores et al. and Yang et al. for a discussion).

Figure 1 shows the distribution of the absolute luminosity ($M_J{\rm{(AB)}}$) of the combined sample. Because in IMAGES we have deliberately selected $M_J{\rm{(AB)}}$$<-20.3$ galaxies, we assume a similar limit for the combined sample.  It let us with a sample of 63 galaxies which is well representative of the luminosity function at z=0.4-0.75. Notice that within this redshift range, GIRAFFE is able to recover the velocity field of almost all galaxies with $W_{0}(\mbox{{\sc[Oii]}})\ge15$\AA. Notice also that the combined sample includes galaxies from 4 independent fields of view (see Yang et al., 2007) and is unlikely affected by cosmic variance effects. In the following we will only discussed the properties of the representative sample of 63 galaxies with $M_J{\rm{(AB)}}$$<-20.3$.

\begin{figure}[!ht]
\begin{center}
\includegraphics[scale = 0.4]{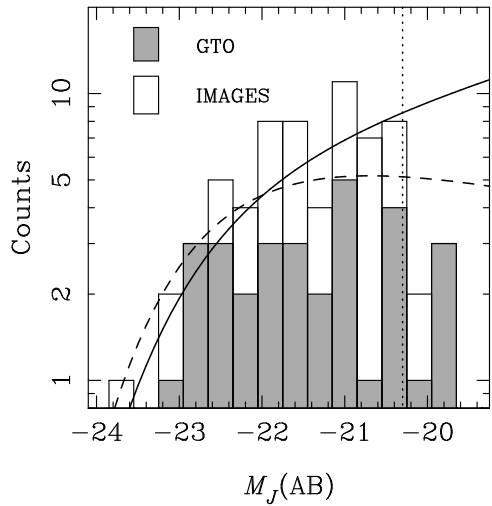}
\end{center}
\caption{Number counts of selected galaxies versus AB absolute
magnitude in $J$-band. The GTO sample refers to Flores et al.\ (2006),
the IMAGES sample to Yang et al. (2007). The vertical
dotted line indicates the limit of the IMAGES program. Two luminosity
functions derived from Pozzetti et al.\ (2003) are shown (full line:
$z\!=\!0.5$; dashed line: $z\,=\,1$). Our sample of galaxies have
redshifts ranging from $z\!=\!0.4$ to $z\!=\!0.75$. It results that
the combined sample of 63 galaxies with $M_J{\rm{(AB)}} \le -20.3$ is
representative of galaxies with stellar masses larger than $1.5\times
10^{10}M_{\odot}$ at $z\sim0.6$.}\label{fig1}
\end{figure}

\subsection{Properties of velocity fields of distant galaxies}
Flores et al. (2006) have generated a classification scheme for the
distant galaxy kinematics. Indeed, at large distances the spatial
resolution is not sufficient to resolve the central regions of the
galaxies. It implies that the observed velocity dispersion ($\sigma$)
of a rotational body is the convolution of the actual random motions
with the rotation. For a rotationally supported galaxy, it unavoidably
leads to a well defined peak in the centre (see Figure 2). Classification of velocity fields and description of their diagnostic may be found in Yang et al. (2007, see also Yang et al. in this volume).
The classification
robustness is provided by the large S/N associated to GIRAFFE
observations and by the fact that large scale motions are well
resolved at z=0.4-0.75, as evidenced by tests on redshifted velocity
fields of various kinds of local galaxies.

Among the 63 velocity fields of the representative sample, Yang et al.
find 20 rotating disks (32\%), 16 rotating disks with perturbations
(25\%) and 27 galaxies with complex kinematics (43\%, see Figure 2).
Within their classification, perturbed rotations correspond to a
discrepant $\sigma$ peak from expectations for a pure rotation.
The large number of galaxies with complex kinematics is even more
puzzling. Indeed, the complex kinematics class corresponds to objects for
which the large scale motions are not aligned to the optical major
axis, and show $\sigma$ map very discrepant from expectations for
a rotation (see Figure 2).

\begin{figure}[t]
\begin{center}
\includegraphics[scale = 0.47, angle=-90]{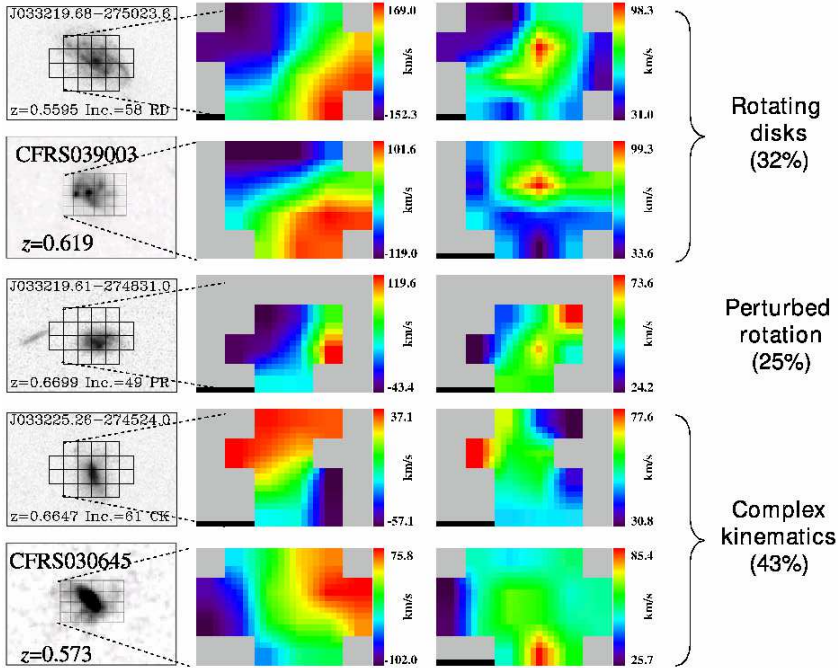}
\end{center}
\caption{Four examples of kinematics of z=0.4-0.75 galaxies; Each row
corresponds to one galaxy, from left to right: HST ACS F755W/F814W
image, observed velocity field and $\sigma$-map. The two top rows show
regular rotating disks with a dynamical axis oriented along the main
optical axis and a $\sigma$-map with a peak dominated by large scale
rotational motions (see text); the three bottom rows show galaxies
with anomalous kinematics, one with just a shift of the peak in the
$\sigma$-map, e.g. a perturbed rotation that may be caused by the
impact of a minor merger, the two other with dynamical axis misaligned
relatively to the main optical axis (the colour version of this figure is available at www.aspbooks.org).}\label{fig1}
\end{figure}

\section{Nature of velocity fields of distant galaxies} 
 We have gathered the largest existing sample of velocity fields for
 distant galaxies. Quite unexpectedly, only a small fraction of
 z=0.4-0.75 galaxies are rotating disks. Let us now consider a
 representative sample of intermediate mass galaxies at z$\sim$0.6. At
 z=0.6, Hammer et al. (1997) found that 60\% of galaxies have
 $W_{0}([OII])\ge15$\AA. Let us assume that quiescent
 ($W_{0}([OII])<15$\AA) galaxy kinematics are either rotationally or
 dispersion supported, thus minimising the fraction of galaxies with
 anomalous kinematics. Then our results imply that 42$\,\pm\,$7\% of
 z=0.4-0.75 galaxies have anomalous kinematics, including 26$\,\pm\,$7\%
 possessing complex kinematics. In the local Universe, such a fraction
 for intermediate mass galaxies has never been thoroughly estimated.
 Because up to 97\% of intermediate mass galaxies are either E, S0 or
 spirals (see Hammer et al., 2005), it is likely that the fraction of
 anomalous kinematics is close to few percents today. 
 
\begin{figure}[!ht]
\begin{center}
\includegraphics[scale = 0.4]{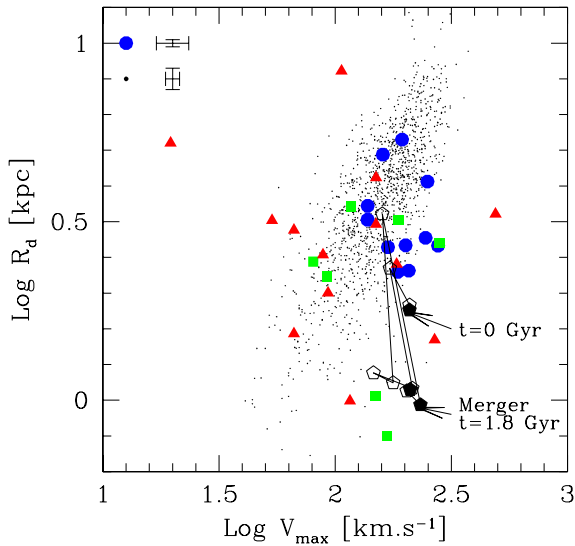}
\end{center}
\caption{Disk scale length versus maximal rotational velocity,
for the sample of 32 distant galaxies of Flores et al. (2006). Blue dots represent galaxies classified as rotating
disks, green squares those classified as perturbed rotators,
and red triangles galaxies with complex kinematics. The small
black dots represent the local sample of Courteau (1997) and
Mathewson et al. (1992). The median 1$\sigma$ uncertainties are
indicated in the upper-left corner, for both distant rotating disks
(blue dots), and local disks (black dots). Black pentagons
represent simulated GIRAFFE observations using a hydrodynamical
simulation of a major merger of two Sbc galaxies.
The beginning and the end of the sequence, as well as the
merger itself, are indicated by solid symbols.}\label{fig1}
\end{figure}
 
 There is a very rapid evolution of galaxy kinematics and it is
unlikely that it may be caused by artificial or instrumental reasons.
Indeed, the observed evolution of the Tully-Fisher relation provides
us with a strong and independent confirmation. By observing galaxies
with slit spectroscopy, Conselice et al. (2005) found a very scattered
Tully-Fisher relation at moderate redshift, as such that one can
wonder whether there is still a correlation between rotational
velocity and stellar mass. Flores et al.\ (2006) and Puech et al.\
(2007) have convincingly shown that galaxies with non-relaxed
kinematics are responsible for the large dispersions of both the
Tully-Fisher and the $j_{\rm disk}$--$V_{\rm max}$ relationships. So
it is beyond doubt that kinematics is among the most rapidly evolving
properties of galaxies. 

Which physical process may explain such a
dramatic evolution ? Puech et al. (2007) have convincingly shown that
merging may be responsible of the anomalous kinematics as well as of
the increase of the scatter of both the Tully-Fisher and the $j_{\rm
disk}$--$V_{\rm max}$ relationships (see also Figure 3). They have
interpreted such a scatter as being due to a random walk process of
galaxies, precisely consistent with encountering between galaxies
during the hierarchical scenario of galaxy formation. Indeed, during a
merger, and especially a major merger, galaxies pass through various
stages during which the disk may be destroyed, generating significant
discrepancies to the general behaviour of isolated rotating disks. A
scenario in which merging is the dominant physical process explains
many evolutionary features since z=1, including the number density
evolution of LIRGs and the emergence of galaxies with complex
morphologies and with blue cores at z$>$ 0.4 (see e.g. Hammer et al.,
2005).

\section{Relation to local galaxies and to disk formation} 
Whether the evolution of galaxy kinematics is associated to (major)
mergers is still a pending question. If true, it means that many
present-day galactic disks have been rebuilt at a relatively recent
epoch. Taking into account the number fraction of both approaching
mergers (those in pairs) and galaxies with complex VFs, Hammer et al.\
(2005, 2007) have concluded that from 50 to 70\% of galaxies may have
experienced a major merger and then a disk rebuilding phase since $z\!=\!1$.
Below we discuss first how this can be consistent with properties of
local galaxies, and then whether other alternatives may explain the
evolution of galaxy kinematics.

Today, intermediate mass galaxies are mostly spiral galaxies (70\% of
them, e.g. Hammer et al., 2005), which are dominated by rotating thin
disks. It is beyond doubt that the Milky Way has not experienced any
major interaction since the last 10-11 Gyrs (z=3). Is this true for
other spiral galaxies or, in other words, how the Milky Way is
representative of local spirals ? Hammer et al. (2007) have compared
the global properties of the Milky Way to those of local spirals with
similar rotational velocities, or $V_{flat}$ which may be taken as a
proxy for the total mass. They show that the Milky Way is quite
discrepant (see Figure 4), showing a disk radius and a stellar content
(or mass) about half that of comparable disk galaxies. Only 7\% of
disk galaxies have properties similar to the Milky Way, while M31 is
quite representative.

Moreover, the distant regions outside of the disk of our Galaxy are
also quite exceptional: compared to those of other galaxies, the stars
inhabiting the outskirts of the Milky Way are exceptionally metal
poor. The galactic outskirt properties may be affected particularly
by past encounters with other galaxies. For example, when two galaxies
merge, tidal debris and newly formed stars produced during the
collision pollute their outskirts. Within this context, outskirts of most spiral galaxies -including M31- show a significant fraction of metal rich stars, evidencing that they have experienced more mergers in their past than did the Milky Way.
All
these pieces of evidence, taken together, support the hypothesis that
the Milky Way is an exceptionally quiet galaxy.

Conversely to the Milky Way, M31 shows evidences for merging in a recent past (e.g. Ibata et al., 2004). 
Galaxy mergers generate additional material in galactic outskirts, including newly formed stars, which then enhance the stellar content and radius. This simply explains the peculiarities of the Milky Way, emphasising that local spirals had a relatively active merger history in the recent past. It is indeed relevant to wonder whether merging may have shaped galaxies -including spirals- as we observe them today.

\begin{figure}[!ht]
\begin{center}
\includegraphics[scale = 0.4]{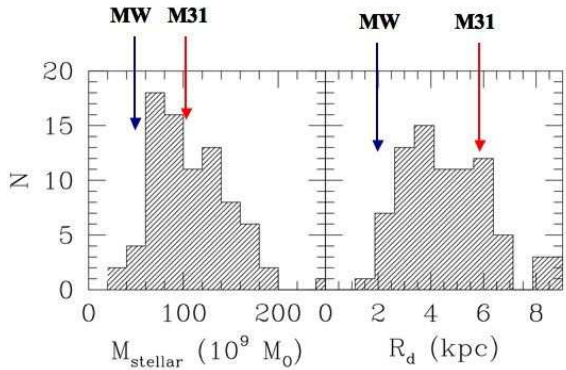}
\end{center}
\caption{Stellar mass (in units of billions solar masses) and disk
radius (in units of kilo-parsecs) distributions of spiral galaxies
with rotational velocities similar to that of the Milky Way. Blue
(red) arrows show the values for the Milky Way (and M31), indicating
that its stellar content and disk radius are half the average values
for external galaxies. To derive these quantities for the Milky Way,
one needs to reconstruct or model the light distribution because we
live within its disk, and extinction by dust may easily hide
significant parts of it. Significant progress has been made recently
(see Hammer et al., 2007), especially using infrared wavelengths to
make such measurements, which are much less affected by dust
extinction.}\label{fig1}
\end{figure}

Although the merger alternative may explain many observations in the
$z\!=\!0.4\!-\!1$ redshift range (e.g., Hammer et al.\ 2005), could a
less dramatic mechanism be at the origin of the peculiar kinematics
found at z=0.4-0.75? In fact, our observations take into account only
the large scale motions of the ionized gas and not those of the
stellar component. While beyond the reach of present-day telescopes,
observations of the latter are certainly crucial, although we still
need to explain why the gas is so turbulent, while more abundant than
it is today. Another possibility may be to advocate for internal
processes such as bars that may perturb the gaseous
velocity fields. Presence of bars in the central region may create
additional dispersion although our spatial resolution is probably too
poor to kinematically characterize bars. Moreover, the absence of
number evolution of barred galaxies (e.g., Zheng et al.\ 2005 and references
therein) does not support that bars could explain the
strong evolution of galaxy kinematics.

\section{Conclusion} 
We do find a strong evolution of the galaxy kinematics since z=0.6, with a significant fraction of galaxies with complex velocity fields. Processes related to merging easily reproduce the observed properties of distant galaxies. Properties of their descendants -the local spiral galaxies- suggest that they have experienced more numerous and recent mergers than the Milky Way. 

However, it has to be demonstrated in details how present-day
$M*$ galaxies, mostly spirals galaxies, can be the end up product of
relatively recent mergers. In fact, galaxy simulations predict that in
most cases, major merger end products are dispersion supported
galaxies, i.e. elliptical galaxies. 
Thus the question of the angular momentum transfer is a key issue. There are indeed considerable uncertainties in the
treatment of the baryonic matter in numerical simulations, which is
illustrated by the so-called angular momentum problem (Steinmetz and
Navarro, 1999), and by the fact that no model has been able to simultaneously
match the luminosity function and Tully-Fisher zero point using
standard $\Lambda$CDM parameters as pointed out by Dutton et al
(2006). To reconcile these discrepancies, Dutton et al. need to assume
an expanding halo caused by a transfer of the angular momentum from a
non-spherical and clumpy gas accretion when forming a primordial disk.
This clearly leads to the question of forming disk by gaseous mergers
as illustrated by Robertson et al. (2006) and by Governato et al.
(2007). The question of the required fraction of gas to form a new disk
after a merger is still pending. It certainly needs some
improvements in the treatment of the angular momentum transfer between
baryonic and dark matter (see Kaufmann et al., 2007).

Observations presented here tell us that major mergers could have played an important role in shaping galaxies as we observe them today. Indeed many estimates of the merger rate found that a typical intermediate mass galaxy should have experienced 0.5 to 0.75 major merger since z= 1 (see Rawat et al, 2007; Kartaltepe et al, 2007 and references therein), and more than 1 since z=2-3. This illustrates further that the Milky Way has had an exceptionally quiet merger history (Hammer et al, 2007). It is unlikely that present-day $M*$ galaxies -mostly spiral galaxies- have all escaped a major merging since z=2-3. Alternatively, one may wonder whether the procedure used to simulate galaxy merger remnants is the correct one.

\acknowledgements 
We are grateful to the technical team at Gepi for the unique achievement of Giraffe at VLT.



\begin{thebibliography}{}
\bibitem[Barnes \& Hernquist(1996)]{1996ApJ...471..115B} Barnes, J.~E., \& Hernquist, L.\ 1996, \apj, 471, 115 
\bibitem[Barton et al.(2000)]{2000ApJ...530..660B} Barton, E.~J., Geller, M.~J., \& Kenyon, S.~J.\ 2000, \apj, 530, 660 
\bibitem[Conselice et al.(2005)]{2005ApJ...628..160C} Conselice, C.~J., Bundy, K., Ellis, R.~S., Brichmann, J., Vogt, N.~P., \& Phillips, A.~C.\ 2005, \apj, 628, 160 
\bibitem[Courteau(1997)]{1997AJ....114.2402C} Courteau, S. 1997, \aj,
  114, 2402
\bibitem[Dressler(2004)]{2004cgpc.symp..206D} Dressler, A.\ 2004, Clusters of Galaxies: Probes of Cosmological Structure and Galaxy Evolution, 206 
\bibitem[Dutton et al.(2006)]{2007ApJ...654...27D} Dutton, A.A., van
  den Bosch, F.C., Dekel, A., et al. 2007, \apj, 654, 27
\bibitem[Flores et al.(1999)]{1999ApJ...517..148F} Flores, H., et al.\ 1999, \apj, 517, 148 
\bibitem[Flores et al.(2006)]{2006A&A...455..107F} Flores, H., Hammer, F., Puech, M., Amram, P., \& Balkowski, C.\ 2006, \aap, 455, 107 
\bibitem[Governato et al.(2007)]{2007MNRAS.374.1479G} Governato, F., Willman, B., Mayer, L., Brooks, A., Stinson, G., Valenzuela, O., Wadsley, J., \& Quinn, T.\ 2007, \mnras, 374, 1479 
\bibitem[Hammer et al.(1997)]{1997ApJ...481...49H} Hammer, F., et al.\ 1997, \apj, 481, 49 
\bibitem[Hammer et al.(2005)]{2005A&A...430..115H} Hammer, F., Flores, H., Elbaz, D., Zheng, X.~Z., Liang, Y.~C., \& Cesarsky, C.\ 2005, \aap, 430, 115 
\bibitem[Hammer et al.(2007)]{2007ApJ...662..322H} Hammer, F., Puech, M., 
Chemin, L., Flores, H., \& Lehnert, M.\ 2007, \apj, 662, 322
\bibitem[Ibata et al.(2004)]{Ibata04} Ibata, R., Chapman, S., Ferguson,
A.~M.~N., Irwin, M., Lewis, G., \& McConnachie, A.\ 2004, \mnras, 351, 117
\bibitem[2007]{kartaltepe2007} Kartaltepe, J. S., Sanders, D. B., et al.
    2007, astro-ph, arXiv:0705.2266
\bibitem[Kaufmann et al.(2007)]{2007MNRAS.375...53K} Kaufmann, T.,
  Mayer, L., Wadsley, J., et al. 2007, \mnras, 375, 53
\bibitem[Ibata et al.(2004)]{2004MNRAS.351..117I} Ibata, R., Chapman,
  S., Ferguson, A.M.N., et al. 2004, \mnras, 351, 117
\bibitem[Le Floc'h et al.(2005)]{2005ApJ...632..169L} Le Floc'h, E.,
  Papovich, C., Dole, H., et al. 2005, \apj, 632, 169
\bibitem[Lilly et al.(1996)]{1996ApJ...460L...1L} Lilly, S.~J., Le Fevre, O., Hammer, F., \& Crampton, D.\ 1996, \apjl, 460, L1 
\bibitem[Mathewson et al.(1992)]{1992ApJS...81..413M} Mathewson, D.S.,
  Ford, V.L., Buchhorn, M. 1992, \apjs, 81, 413
    \bibitem[Pozzetti et al.(2003)]{2003A&A...402..837P} Pozzetti, L., et al.\ 2003, \aap, 402, 837 
\bibitem[Puech et al.(2007)]{2007A&A...466...83P} Puech, M., Hammer, F., Lehnert, M.~D., \& Flores, H.\ 2007a, \aap, 466, 83 
\bibitem[Rawat et al.(2007)]{} Rawat, A., Hammer, F., Kembhavi, A.~K., \& Flores, H., \ 2007, \apjl, submitted 
\bibitem[Robertson et al.(2006)]{2006ApJ...645..986R} Robertson, B., Bullock, J.~S., Cox, T.~J., Di Matteo, T., Hernquist, L., Springel, V., \& Yoshida, N.\ 2006, \apj, 645, 986 
\bibitem[Steinmetz \& Navarro(1999)]{1999ApJ...513..555S} Steinmetz,
  M. \& Navarro, J.F. 1999, \apj, 513, 555
  \bibitem[Yang et al.(2007)]{} Yang, Y. B., Flores, H., Hammer, F. et al.  \ 2007, \aap, submitted 
\bibitem[Zheng et al.(2005)]{2005A&A...435..507Z} Zheng, X.~Z., Hammer, F., Flores, H., Ass{\'e}mat, F., \& Rawat, A.\ 2005, \aap, 435, 507 
\end{thebibliography}
\end{document}